\documentclass[%
 aip,
 amsmath,amssymb,
 reprint,%
]{revtex4-1}

\usepackage{caption}

\usepackage{titlesec}
\titlespacing{\section}
  {-5pt}{1.\baselineskip}{0.5\baselineskip}
\usepackage{graphicx}
\usepackage{dcolumn}
\usepackage{bm}

\usepackage[utf8]{inputenc}
\usepackage[T1]{fontenc}
\usepackage{mathptmx}
\begin{document}

\preprint{AIP/123-QED}

\title[Spin pumping and inverse spin Hall effect in iridium oxide]{Spin pumping and inverse spin Hall effect in iridium oxide}

\author{Biswajit Sahoo}
\author{Koustuv Roy}%
\author{Pushpendra Gupta}
\affiliation{%
	Laboratory for Nanomagnetism and Magnetic Materials (LNMM), School of Physical Sciences; National Institute of Science Education and Research (NISER), HBNI, Jatni-752050, India.\\
}%
\author{Biswarup Satpati}
\affiliation{ 
Saha Institute of Nuclear Physics, 1/AF Bidhannagar,  Kolkata 700064, India
}%
\author{Braj Bhusan Singh}
\author{Subhankar Bedanta}
\email{sbedanta@niser.ac.in}
\affiliation{%
	Laboratory for Nanomagnetism and Magnetic Materials (LNMM), School of Physical Sciences; National Institute of Science Education and Research (NISER), HBNI, Jatni-752050, India.\\
}%

\date{\today}

\begin{abstract}
Large charge-to-spin conversion (spin Hall angle) and spin Hall conductivity are prerequisites for development of next generation power efficient spintronic devices. In this context, heavy metals (e.g. Pt, W etc.), topological insulators, antiferromagnets are usually considered because they exhibit high spin-orbit coupling (SOC). In addition to the above materials, 5d transition metal oxide e.g. Iridium Oxide (IrO$_2$) is a potential candidate which exhibits high SOC strength.  Here we report a study of spin pumping and inverse spin Hall effect (ISHE), via ferromagnetic resonance (FMR), in IrO$_2$/CoFeB system. We identify the individual contribution of spin pumping and other spin rectification effects in the magnetic layer, by investigating the in-plane angular dependence of ISHE signal. Our analysis shows significant contribution of spin pumping effect to the ISHE signal. We show that polycrystalline IrO$_2$ thin film exhibits high spin Hall conductivity and spin Hall angle which are comparable to the values of Pt.
\end{abstract}

\maketitle

The field of ``spintronics'', where both charge and spin degrees of freedom are exploited, has been a topic of active research since the past few decades. With charge based electronics reaching their limitations, more attention is now being given to devices and materials which employ spin currents. A radical shift from the conventional electronic paradigm, devices based on spintronics have shown potential application in data storage, non-volatile magnetic random access memories (MRAMs) \cite{Kimura2018,Zhang2018,Hu2011}, spin based field effect transistors \cite{Dankert2017} etc. In such devices, spin-orbit coupling (SOC) plays a vital role. SOC is a relativistic phenomena which couples the spin angular momentum to the orbital angular momentum of an electron. SOC is responsible for both spin Hall effect (SHE) and inverse spin Hall effect (ISHE) \cite{Sinova2015,Hirsch1999,Liu2012} i.e, it can convert charge current to spin current and \textit{vice versa} respectively. 

Spin pumping is the phenomenon of transfer of spin angular momentum from a ferromagnet (FM) to an adjacent non-magnetic (NM) material in the presence of an external magnetic field and an excitation microwave field \cite{Sinova2015}. It should be noted that the NM should have high SOC for pronounced SHE and ISHE. The precessing magnetic moments in the FM pump pure spin current into the NM layer which is subsequently dissipated by various relaxation processes. The magnitude of this spin current is maximum under the condition of ferromagnetic resonance. The spin current has the following form \cite{Tserkovnyak2005}:

\begin{equation}
	J_s=\frac{\hbar}{4\pi}g^{\uparrow\downarrow}_{eff}\hat{m}\times\frac{d\hat{m}}{dt}
\end{equation}

Here $g^{\uparrow\downarrow}_{eff}$ is the effective spin mixing conductance and $\hat{m}$ is the magnetization unit vector. If the NM has a high SOC, then due to ISHE, the spin current may be detected electrically. As spin pumping opens up an additional channel of angular momentum dissipation, the intrinsic damping of the FM increases \cite{Tserkovnyak2002} which is given as:

\begin{equation}
	\Delta\alpha=\frac{g\mu_B}{4\pi M_sd_{FM}}g^{\uparrow\downarrow}_{eff}
	\label{spin mix}
\end{equation} 
Here $\Delta \alpha$ is the difference between the Gilbert damping of the FM/NM bilayer and the reference FM single layer. $M_s$ is the saturation magnetization of the bilayer, g is the gyromagnetic ratio, $\mu_B$ is the Bohr magneton and $d_{FM}$ is the ferromagnetic thickness. $g^{\uparrow\downarrow}_{eff}$ denotes the effective spin mixing conductance, an important parameter that characterizes the efficiency of transfer of spins through the FM/NM interface.
However, damping may also increase due to inhomogeneities in the sample, interfacial defects, magnetic proximity effects, presence of capping layer \cite{Conca2016,Ruiz-Calaforra2015} etc. Therefore, only increase in damping cannot be taken as a sure indication of spin-pumping in a system \cite{Conca2017}. This is where electrical measurements for detection of spin pumping become necessary.

 As SOC is proportional to Z$^4$ (Z being the atomic number), usually heavy metals such as Pt (Z=78) \cite{Rojas-Sanchez2014}, Pd (Z=46)  \cite{Ando2010}, W (Z=74) \cite{Pai2012}, Ta (Z=73) \cite{Kim2015} etc, have been used for spin current detection via ISHE. However, it has been shown that 5d transition metal oxides, such as iridium oxide, also exhibits high SOC, which originates due to the 5d electrons in the conduction band \cite{Fujiwara2013}.  The expectation value of the SOC ($\langle L.S \rangle$) is $\approx$ 1.0$\hbar^2$ for Ir (Z=77), while it is $\approx$3.1$\hbar^2$ for IrO$_2$ \cite{Clancy2012}. This shows that there is a significant SOC in iridium oxide. Further, Fujiwara \textit{et. al.} \cite{Fujiwara2013}, have shown successful injection of spin current in IrO$_2$ wires in lateral spin valve geometry. However, spin pumping via ferromagnetic resonance (FMR) in iridium oxide/ ferromagnet bilayer thin films has not been explored till now. Further we have estimated the spin mixing conductance for iridium oxide.

The thin films have been prepared on Si p-(100) substrates with native oxide, in a high vacuum chamber (manufactured by Mantis Deposition Ltd. UK) with base pressure better than 1.1 $\times10^{-7}$ mbar. We name the samples as S1, S2, S3, S4 and S5 for Si/Co$_{40}$Fe$_{40}$B$_{20}$(10), Si/Co$_{40}$Fe$_{40}$B$_{20}$(10)/IrO$_2$(2,3,5) and Si/Co$_{40}$Fe$_{40}$B$_{20}$(10)/Ir(3)/Cr(3), respectively. The numbers in brackets are in nm. Here S5 is the control sample.  Co$_{40}$Fe$_{40}$B$_{20}$ (CoFeB) and Ir were deposited via DC magnetron sputtering. For  CoFeB the deposition pressure was $\sim $7 $\times 10^{-4}$ mbar. The substrate was rotated at 20 rpm while deposition of all the layers, in order to provide sample uniformity. The deposition rate was 0.021 nm/s. For deposition of iridium oxide (samples S2-S4), we followed the below procedure\cite{Liao1998}. First, we increased the temperature of the substrate to 150$^\circ$C. Then we pumped oxygen gas into the chamber. The total pressure of Ar and O$_2$ was kept constant at 2.5 $\times10^{-3}$ mbar. Then Ir was DC sputtered from a 99.9\% pure Ir target, at a rate of 0.016 nm/s onto the substrate. For Sample S5, CoFeB and Ir were DC sputtered in the absence of oxygen while e-beam evaporation was used for Cr deposition. The deposition rates of Ir and CoFeB was 0.013 nm/s and 0.021 nm/s, respectively. CoFeB was deposited at room temperature while Ir and Cr were deposited at 150$^0$C.

For x-ray diffraction (XRD) we have deposited a 50 nm thick IrO$_2$ film on Si p-(100) substrate. As shown in Figure S1, we find a peak at $2\theta= 35^\circ $, corresponding to IrO$_2$ (101) crystal \cite{Kim2008}. For samples S1- S5 we have also performed x-ray reflectivity (XRR) to evaluate the thickness of the layers and the interface roughness. The XRD and XRR measurements have been performed with a Rigaku SmartLab diffractometer. Further for checking the interface and
crystalline quality we have performed cross-sectional transmission electron microscopy (TEM). Images were acquired by 300 kV TEM system of FEI, Tecnai G2 F30, S-Twin microscope, which was equipped with Gatan Orius CCD camera, HAADF detector, scanning unit and EDX spectrometer. We have performed energy dispersive X-Ray (EDX) spectroscopy on the samples to observe the growth of the various layers and interface quality (intermixing).

\begin{figure}[h!]
	\centering
	\includegraphics[scale=0.3]{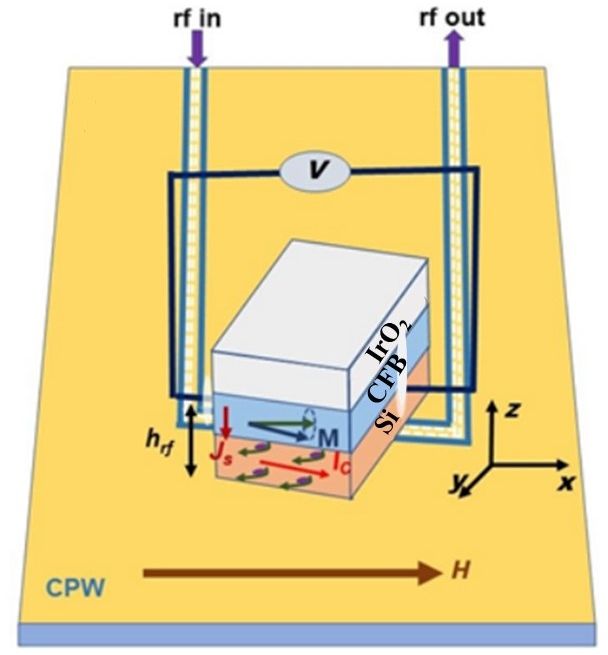}
	\caption{Experimental set-up for measurement of ISHE signal in the sample structure Si/CoFeB(10)/IrO$_2$(x).}
	\label{fig:setup}
\end{figure}

We performed frequency dependent FMR measurements (Phase FMR manufactured by NanoOsc, Sweden) on these samples to determine the intrinsic Gilbert damping. We have used a coplanar waveguide FMR (CPW-FMR) set up, the schematics of which is shown in Figure \ref{fig:setup}. The sample is placed face down on the waveguide, which transmits a radio frequency electromagnetic (EM) wave through it. We used a modified FMR setup to detect ISHE in these samples, the details of which are described elsewhere  \cite{Singh2017,Singh2018,Singh2019,Singh2020,gupta2020}. Further, we performed in-plane angular dependent measurement of the voltage with respect to the external magnetic field, to disentangle the contribution of spin pumping and other spin rectification effects.

\begin{figure*}[t]
	\centering
	\includegraphics[width=1\textwidth]{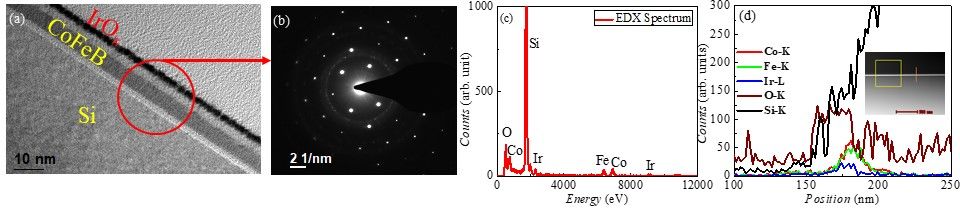}
	\caption{a) High resolution XTEM image of Si/CoFeB(10)/IrO$_2$ (2nm).  (b) Selected area diffraction pattern from a region shown in (a). (c) EDX spectrum showing the presence of all the elements in the sample except B. (d) EDX line scan along the line shown in the STEM-HAADF image (inset). There is a clear layer of IrO$_2$ formed with an additional layer of possibly oxide of CoFeB with inter-dispersed Ir.}
	\label{fig:TEM}
\end{figure*}

The TEM image in Figure \ref{fig:TEM}(a) shows a formation of about 2 nm thick layer of poly-crystalline IrO$_2$ in sample S2. There has been oxidation of CoFeB, as evidenced by an extra layer in between the IrO$_2$ and CoFeB layer. The EDX measurements show definite presence of oxygen and iridium in the sample, hinting significant formation of IrO$_2$. The ratio of Ir to oxygen is not 1:2, suggesting presence of pure Ir as well. EDX line scan using scanning TEM (STEM) high angle annular dark-field (HAADF) mode shows the diffusion of Ir into the oxide layer.

\begin{figure}[h!]
	\centering
	\includegraphics[scale=0.25]{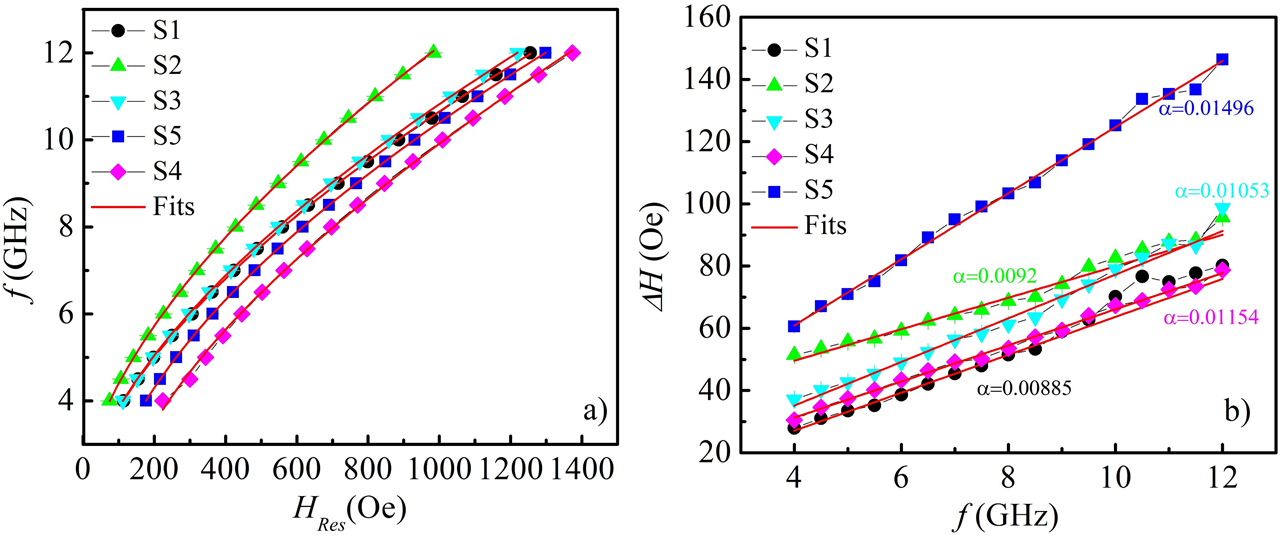}
	\caption{a) shows the Kittel fitting for all the samples. The red lines indicate best fits. b) shows the $\Delta$H vs frequency dependence. The red lines indicate best fits.}
	\label{fig:KitDamp}
\end{figure}

FMR spectra were recorded over a frequency ($f$) range of 4 to 12 GHz, with an interval of 0.5 GHz for all the three samples. The FMR spectra are shown in Fig. S2 of supplementary information. The spectra was fitted to a Lorentzian function to obtain the resonance field (\textit{H$_{Res}$}), and the linewidth (\textit{$\Delta$H}). 
We fitted the $f$-$H_{Res}$ data shown in Fig. \ref{fig:KitDamp}(a) to the following Kittel equation   \cite{Kittel1947}:

\begin{equation}
	f=\frac{\gamma}{2\pi}\sqrt{(H_{Res}+H_k)(H_{Res}+H_k+4 \pi M_{eff})}
	\label{Kittel}
\end{equation}

Here $H_k$ is the uniaxial anisotropy field, $\gamma$ is the gyromagnetic ratio and $M_{eff}$ is the effective demagnetization.

We used the value of $\gamma$ obtained from Equation \ref{Kittel} to fit the $\Delta$$H$- $f$ data shown in Fig \ref{fig:KitDamp}(b) to obtain the 
Gilbert damping $\alpha$ by the following equation:

\begin{equation}
	\Delta H=\frac{4\pi \alpha}{\gamma}f+\Delta H_0
\end{equation}

where $\Delta H_0$ is the inhomogeneous linewidth broadening.

From the best fits (Figure \ref{fig:KitDamp}b), we found the values of $\alpha$ of samples S1-S5 to be $\approx$ 0.0088$\pm 0.0002$, $0.0092\pm0.0002$, $0.0105 \pm 0.0001$ ,$0.0115 \pm 0.0001$ and $0.0150\pm0.0001$, respectively.  The enhancement of $\alpha$ for S2-S4 in comparison to S1 indicates the presence of spin pumping. However, other interfacial effects may also contribute to the enhancement of $\alpha$.   \cite{Conca2017} 

\begin{figure}[h!]
	$\begin{array}{rl}
	\multicolumn{1}{c}{\includegraphics[width=0.5\textwidth]{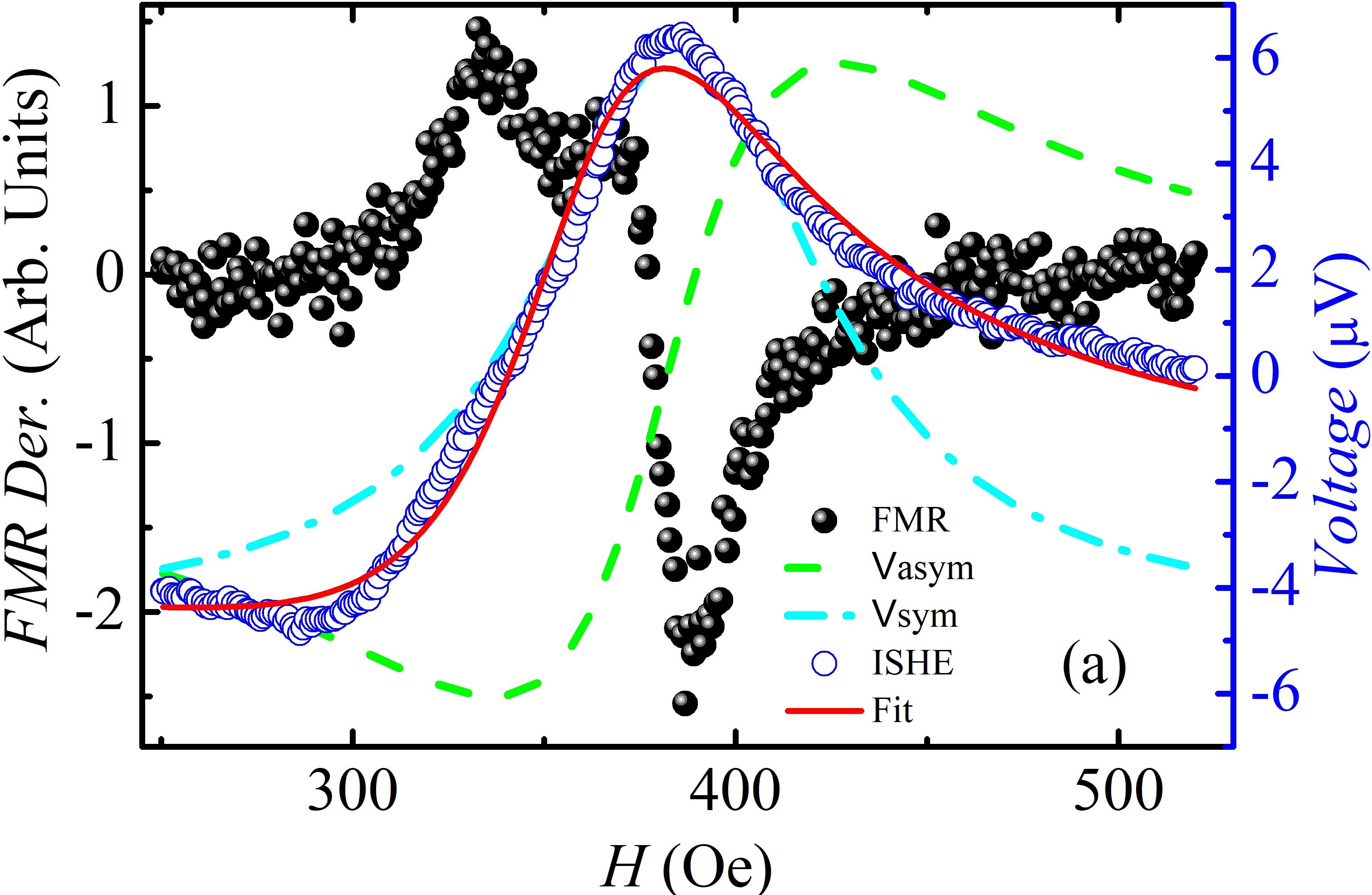}}\\
	\includegraphics[width=0.5\textwidth]{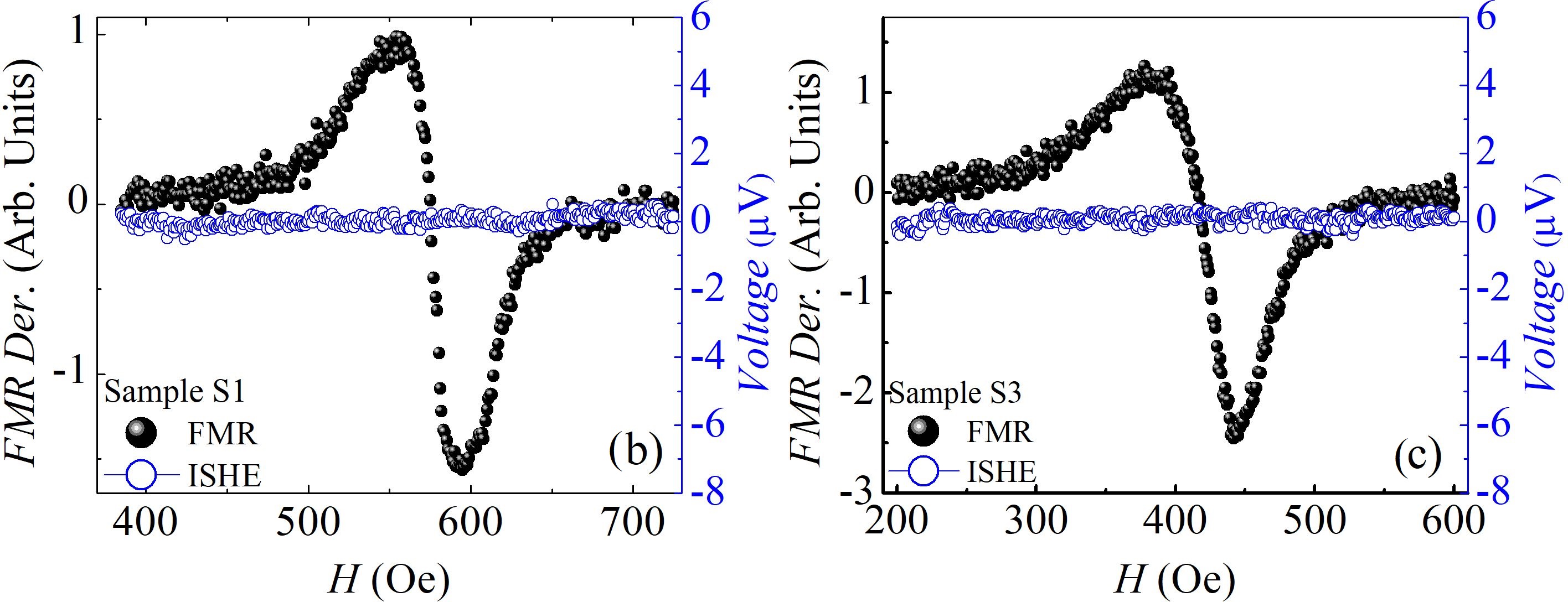}
	
\end{array}$

\caption{ISHE signals for samples a) S2 b)S1 and c) S5. We clearly see a definite contribution of symmetric and a anti-symmetric signals to ISHE in case of S2. S1 and S3 on the other hand show a dominant noise contribution. The r.f power used in this case is $\sim$50mW at an r.f. frequency is 7 GHz.}
\label{fig:ISHE}
\end{figure}

In order to confirm that the enhancement of $\alpha$ is primarily due to spin pumping, we have performed ISHE measurements on all the samples. We have observed ISHE voltage in samples S2-S4. Figure \ref{fig:ISHE}(a) shows the ISHE signal (open circles) and the corresponding FMR signal (solid circles) in S2. While the pure ISHE signal is symmetric in nature, spin rectification effects from the ferromagnetic layer have both symmetric ($V_{sym}$) and anti-symmetric ($V_{asym}$) components  \cite{Harder2016,Soh2014,Gui2012}. Therefore we used the following Lorentzian function to separate the symmetric and anti-symmetric components from the measured voltage ($V_{meas}$) . 

\begin{flalign} 
	V_{\text { meas }}=& V_{\text { sym }} \frac{(\Delta H)^{2}}{\left(H-H_{\mathrm{Res}}\right)^{2}+(\Delta H)^{2}}\notag \\ &+V_{\text { asym }} \frac{ \Delta H\left(H-H_{\mathrm{Res}}\right)}{\left(H-H_{\mathrm{Res}}\right)^{2}+(\Delta H)^{2}} 
	\label{eq:fit}
\end{flalign} 

The green and cyan dotted lines in Figure \ref{fig:ISHE}(a) show the symmetric and anti-symmetric components respectively, while the red line shows the overall fit to Equation \ref{eq:fit}. It is to be noted that we could not observe any measurable ISHE voltage for samples S1 and S3 as shown in Figure \ref{fig:ISHE}(b) and (c), respectively. This indicates that the presence of IrO$_2$ in sample S2 is imperative to produce measurable spin pumping/ISHE.

In-plane angular dependent voltage measurements were performed in order to quantify the contribution of different spin rectification effects like anomalous Hall effect (AHE) and anisotropic magneto-resistance (AMR) to the obtained voltage signal. From the spin pumping effect, we have performed in-plane angle dependent measurement of the $V_{meas}$ at $f$ = 7 GHz at 5$^\circ$ intervals. The angle $\phi$ is defined as the angle between the h$_{rf}$ and the external magnetic field, which is 90$^\circ$ in our case.

\begin{figure}[htbp]
	\centering
	\includegraphics[width=1\linewidth]{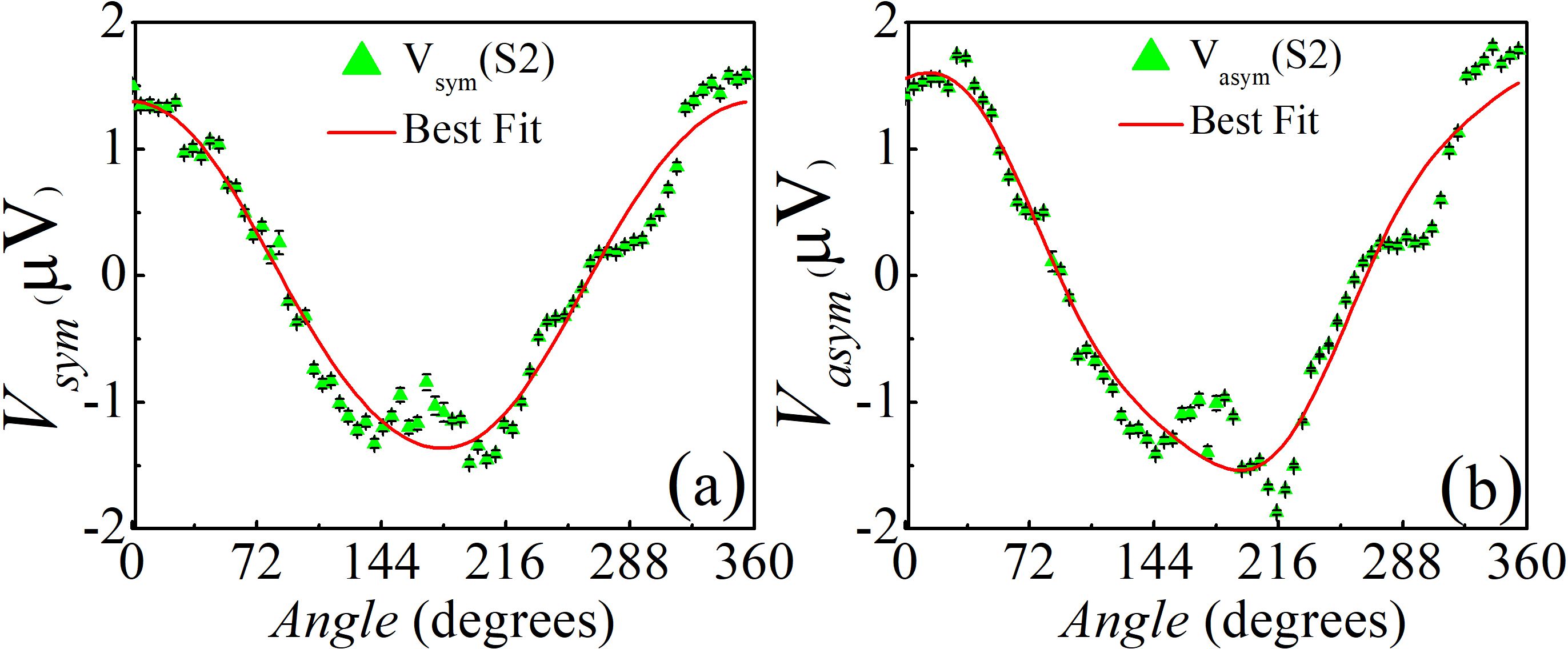}
	\caption{In-plane angular dependence of obtained voltage with respect to the external magnetic field. The plots are for the symmetric components extracted from the overall signal as shown in Figure \ref{fig:ISHE} The best fit lines are obtained as per Equations \ref{eq:Sym} and \ref{eq:Anti-Sym}.  }
	\label{fig:anglesym}
\end{figure}

We plot the symmetric and anti-symmetric components vs $\Phi$ for samples S2-S4 (figure \ref{fig:anglesym} and supplementary figure S3). Using the model developed by Harder \textit{et al}.  \cite{Harder2016}, we then quantify the contribution of spin pumping and various spin rectification effects in the samples . The equations used are:

\begin{eqnarray}
\begin{aligned} V_{\mathrm{sym}}=& V_{\mathrm{sp}} \cos ^{3}(\Phi)+V_{\mathrm{AHE}} \cos (\phi) \cos (\Phi)\\&+V_{\mathrm{AMR}\text{-}\perp}^{\mathrm{sym}} \cos (2 \Phi) \cos (\Phi) \\ &+V_{\mathrm{AMR}\text{-}\|}^{\mathrm{sym}} \sin (2 \Phi) \cos (\Phi) \\
	\label{eq:Sym}
V_{\text {asym}}=& V_{\text {AHE}} \sin (\phi) \cos (\Phi)+V_{\text {AMR}\text{-}\perp}^{\text {asym}} \cos (2 \Phi) \cos (\Phi) \\ &+V_{\text {AMR}\text{-}\|}^{\text {asym}} \sin (2 \Phi) \cos (\Phi) \end{aligned}\\
\label{eq:Anti-Sym}
\end{eqnarray}

where $\Phi$ is the angle between the line of contacts on the sample and applied magnetic field. 

From the fitted plots, we see that for sample S2, there is nearly equivalent contribution from both symmetric and anti-symmetric components. At a r.f. power of $\approx$ 11mW, the spin pumping ($V_{SP}$) was found to be 2.79 $\pm$ 0.09 $\mu V$. The AHE voltage ($V_{AHE}$) was 1.64 $\pm $ 0.06 $\mu V$. The AMR contribution is calculated as: $V_{\text{AMR-}(\|,\perp)}=\sqrt{\big(V^{\text{ASym}}_{\text{AMR-}(\|,\perp)}\big)^2+\big(V^{\text{Sym}}_{\text{AMR-}(\|,\perp)}\big)^2}$. For sample S2, $V_{\text{AMR-}\|}=0.23\pm 0.10 \mu V$ and $V_{\text{AMR-}\perp}=1.2\pm 0.08 \mu V$. The value of  $V_{\text{AMR-}\|}$ is one order less than the $V_{AHE}$ or $V_{SP}$.  $V_{\text{AMR-}\perp}$ has a significant contribution to the voltage signal. However, the model could encapsulate the observed behavior satisfactorily with minimal deviations. Similarly, for S3 and S4, we obtain $V_{SP}=2.65\pm 0.02$$\mu V$ and 1.55 $\pm 0.01$$\mu V$ and $V_{AHE}=1.62 \pm 0.01 \mu V$ and $0.361 \pm0.002\mu V$ respectively. In all the three samples, $V_{SP}$ had the highest contribution to the ISHE signal. 

We have calculated the spin mixing conductance ($g^{\uparrow\downarrow}_{eff}$) using the following equation \cite{Tserkovnyak2002,Ando2011,Mosendz2010}, in order to determine the efficiency of transfer of spins through the FM/NM interface. 
\begin{equation}
g^{\uparrow\downarrow}_{eff}=\frac{4\pi M_sd_{FM}\Delta\alpha}{g\mu_B}
\label{Eq:spin mixing conductance}
\end{equation}
This was then used in evaluation of the spin Hall angle (SHA) of the three samples. The saturation magnetization M$_s$ was measured by a SQUID magnetometer. For S2-S4, it is $\approx$ 704.32 $\pm 2.12$ emu/cc, 665.68 $\pm 3.20$ emu/cc and 696.59 $\pm 1.45$ emu/cc, respectively . From Equation \ref{Eq:spin mixing conductance}, the $g_{eff}^{\uparrow\downarrow}$ for samples S2-S4 are evaluated to be 0.157$\pm 0.001 \times 10^{19}$ m$^{-2}$, 0.755$\pm 0.007 \times 10^{19}$ m$^{-2}$ and 0.915$\pm0.008 \times 10^{19}$ m$^{-2}$, respectively. 

Further, we have calculated the spin Hall angle ($\theta_{SHA}$) for samples S2-S4. $\theta_{SHA}$ is the figure of merit of any spintronic device and shows the ratio of conversion of spin current to charge current.  $\theta_{SHA}$ is related to the $V_{ISHE}$ by the following relation  \cite{Rogdakis2019}

\begin{equation}
V_{I S H E}=\left(\frac{w}{\frac{t_{IrO_2}}{\rho_{IrO_2}}+\frac{t_{CFB}}{\rho_{CFB}}}\right)\times \theta_{S H A} \lambda_{S D} \tanh \left[\frac{t_{P t}}{2 \lambda_{S D}}\right] J_{s}
\end{equation}

where $J_s$ is given as:
\begin{eqnarray}
\begin{aligned}[left]
	J_{s} &\approx\left(\frac{g_{r}^{\uparrow \downarrow} \hbar}{8 \pi}\right)\left(\frac{\mu_{0} h_{r f} \gamma}{\alpha}\right)^{2} \times \\& \left[\frac{\mu_{0} M_{s} \gamma+\sqrt{\left(\mu_{0} M_{s} \gamma\right)^{2}+16(\pi f)^{2}}}{\left(\mu_{0} M_{s} \gamma\right)^{2}+16(\pi f)^{2}}\right]\left(\frac{2 e}{\hbar}\right)
\end{aligned}
\end{eqnarray}

$g_r^{\uparrow\downarrow}$ is the real part of $g_{eff}^{\uparrow\downarrow}$ (Equation \ref{Eq:spin mixing conductance}) and  is given as:
\begin{equation}g_{r}^{\uparrow \downarrow}=g_{e f f}^{\uparrow \downarrow}\left[1+\frac{g_{e f f}^{\uparrow \downarrow} \rho_{IrO_2} \lambda_{IrO_2} e^{2}}{2 \pi \hbar \tanh \left[\frac{t_{IrO_2}}{\lambda_{IrO_2}}\right]}\right]^{-1}\end{equation}

With $\rho_{IrO_2}=100.5\pm0.9 \mu\Omega cm$ and $\rho_{CFB}=700.1\pm0.9 \mu\Omega cm$, the $\theta_{SHA}$ for S2-S4 is found to be $\sim$ 0.19 $\pm 0.02$, 0.05$\pm 0.01$ and 0.02$\pm 0.01$, respectively. The value of SHA decreases with increasing thickness of IrO$_2$. As the spin diffusion length of IrO$_2$ is 3.8 nm $\cite{Fujiwara2013}$, more spin relaxation occurs as we increase the thickness of IrO$_2$. This results in lowering of the ISHE signal with increasing thickness, and correspondingly less SHA value for thicker samples.

The spin Hall conductivity is given as $\sigma_{SH}= \frac{\theta_{SHA}}{\rho_C} \times \frac{\hbar}{2e}$. In our case, the values obtained are 1.93$\pm0.21 \times 10^5\frac{\hbar}{2e}\Omega^{-1}m^{-1}$, 0.54$\pm 0.12\times 10^5\frac{\hbar}{2e}\Omega^{-1}m^{-1}$ and 0.20$\pm0.04\times 10^5\frac{\hbar}{2e}\Omega^{-1}m^{-1}$ for samples S2-S4, respectively. The $\sigma_{SH}$ values obtained for IrO$_{2}$ 2nm is higher as compared to many other high SOC materials like Pt, W, Ta,  Bi$_{2}$Se$_{3}$ etc. \cite{Khang2018}.

To summarize, we have successfully fabricated samples of IrO$_2$/CoFeB bilayers and have confirmed the same by XRD and TEM measurements. The IrO$_2$ formed is polycrystalline in nature. We could get ISHE signals with r.f power as low as 6 mW. This shows that the present system is highly receptive to low magnitude of spin current. Occurrence of spin pumping was indicated by enhancement of $\alpha$ and was subsequently quantified by in-plane angle dependent ISHE experiments. Spin pumping was shown to have the dominant contribution. We further determine the $\theta_{SHA}$ ranging from $\approx$ 0.19 to 0.02 depending on the thickness. The largest spin Hall conductivity was determined to be $1.93 \times 10^{5}$ $\Omega^{-1}m^{-1}$ for IrO$_2$ of 2 nm thickness. The high value of $\theta_{SHA}$ may be due to back-scattering and reflection of spins at the interface. Further experimental and theoretical efforts are required to understand the occurrence of the high spin Hall angle in IrO$_2$. Moreover, due to presence of Dirac Nodal Lines (DNLs) \cite{Nelson2019,Xu2019} in epitaxial IrO$_2$ thin films, one might be able to tune the $\sigma_{SH}$ to high ( or low) values which is suitable for a spin current generator (or detector). The possible tunability of the $\sigma_{SH}$ in IrO$_2$ warrants further investigation for its applicability in devices and thin-films.

\begin{acknowledgments}
We thank the Department of Atomic Energy for providing financial support. BBS acknowledges DST for INSPIRE faculty fellowship. PG and KR acknowledge UGC and CSIR for JRF fellowships, respectively.
\end{acknowledgments}

\section*{Data Availability Statement}
The data that support the findings of this study are available from the corresponding author upon reasonable request.

\nocite{*}
\bibliography{aipsamp}

\end{document}


\date{}
	
\title{Supplementary Information: \\Spin pumping and inverse spin Hall effect in iridium oxide}
	
\author[1]{Biswajit Sahoo}
\author[1]{Koustuv Roy}
\author[1]{Pushpendra Gupta}
\author[2]{Biswarup Satapati}
\author[1]{Braj Bhusan Singh}
\author[1]{Subhankar Bedanta\thanks{sbedanta@niser.ac.in}}
\affil[1]{Laboratory for Nanomagnetism and Magnetic Materials (LNMM)\\
	School of Physical Sciences\\
	National Institute of Science Education and Research (NISER)\\
	HBNI, Jatni-752050, India.}
\affil[2]{Saha Institute of Nuclear Physics, 1/AF Bidhannagar,  Kolkata 700064, India}

	
	\maketitle
\section{X-Ray diffraction (XRD)}
\begin{figure}[h!]
	\centering
	\includegraphics[width=0.6\linewidth]{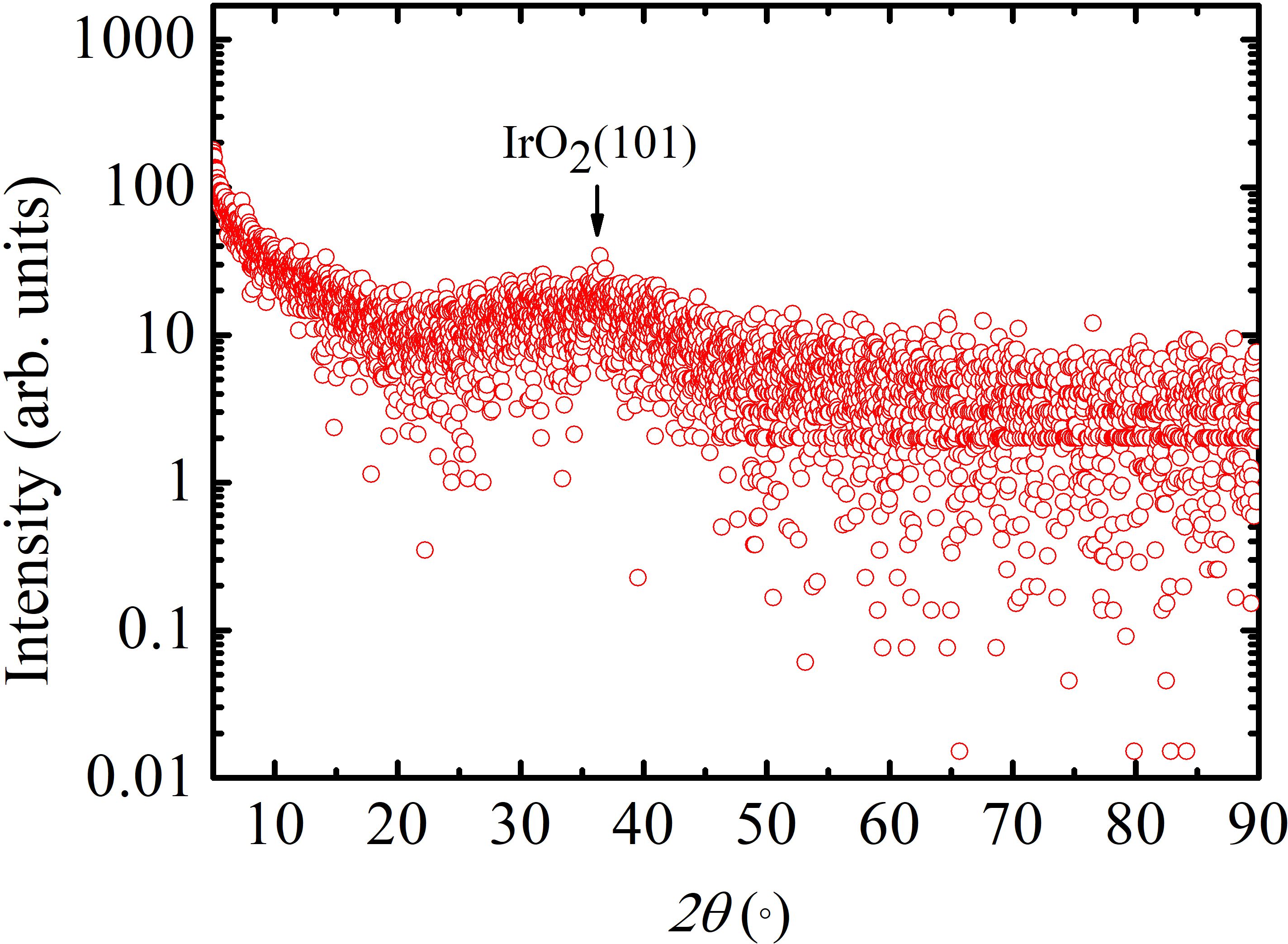}
	\caption{XRD of 50nm IrO$_2$ single layer thin film.}
	\label{fig:XRD}
\end{figure}

Fig \ref{fig:XRD} shows the XRD data for the single layer of IrO$_2$ film on Si p-(100) substrate. A weak peak was observed at around 2$\theta=35^\circ$ corresponding to IrO$_2$ (101) crystal  (Kim \textit{et al.} Journal of Applied Physics \textbf{103}, 023517 (2008)). 
	
\section{Ferromagnetic Resonance}

\begin{figure*}[t]
	\centering
	\includegraphics[width=.315\textwidth]{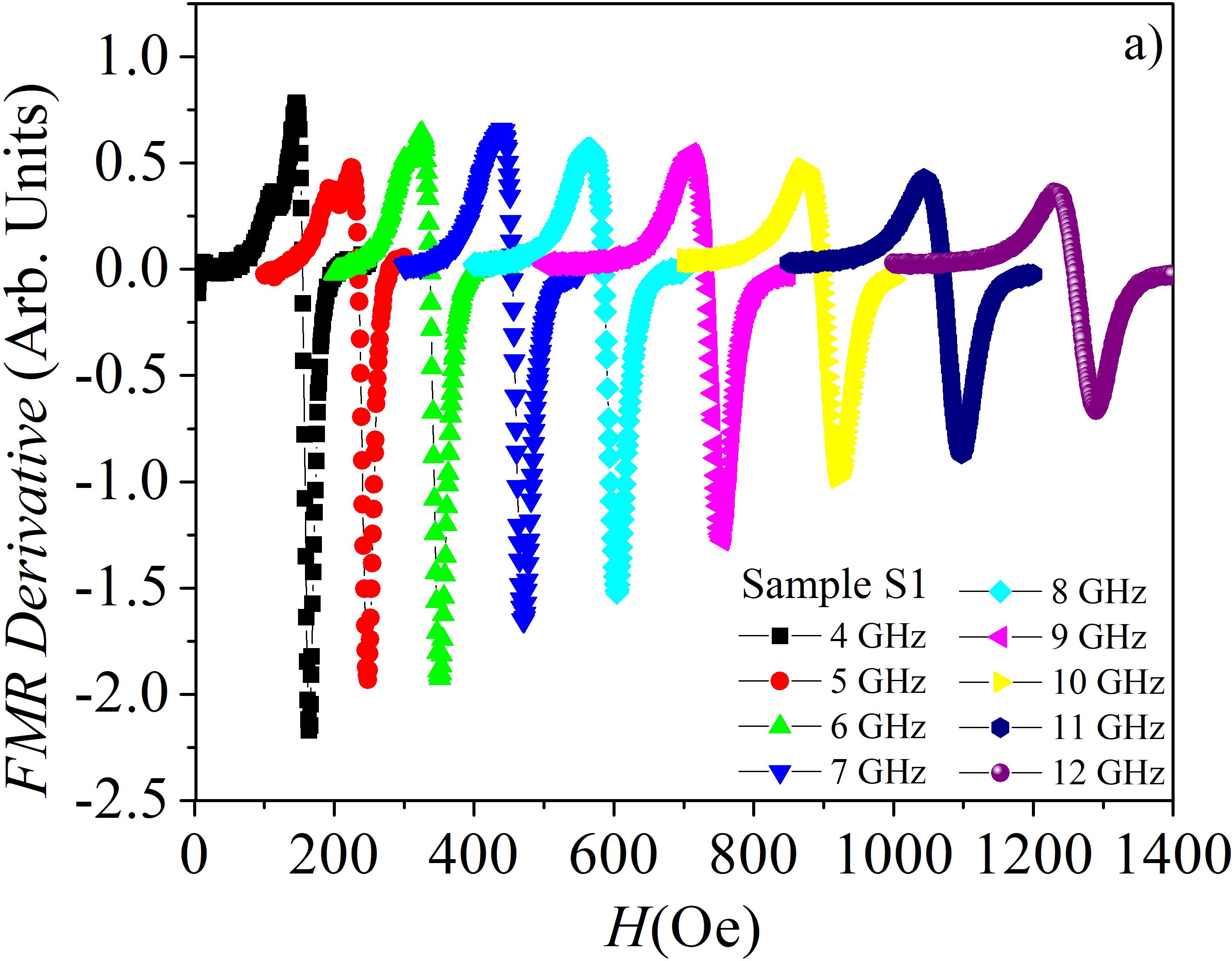}\hfill
	\includegraphics[width=.315\textwidth]{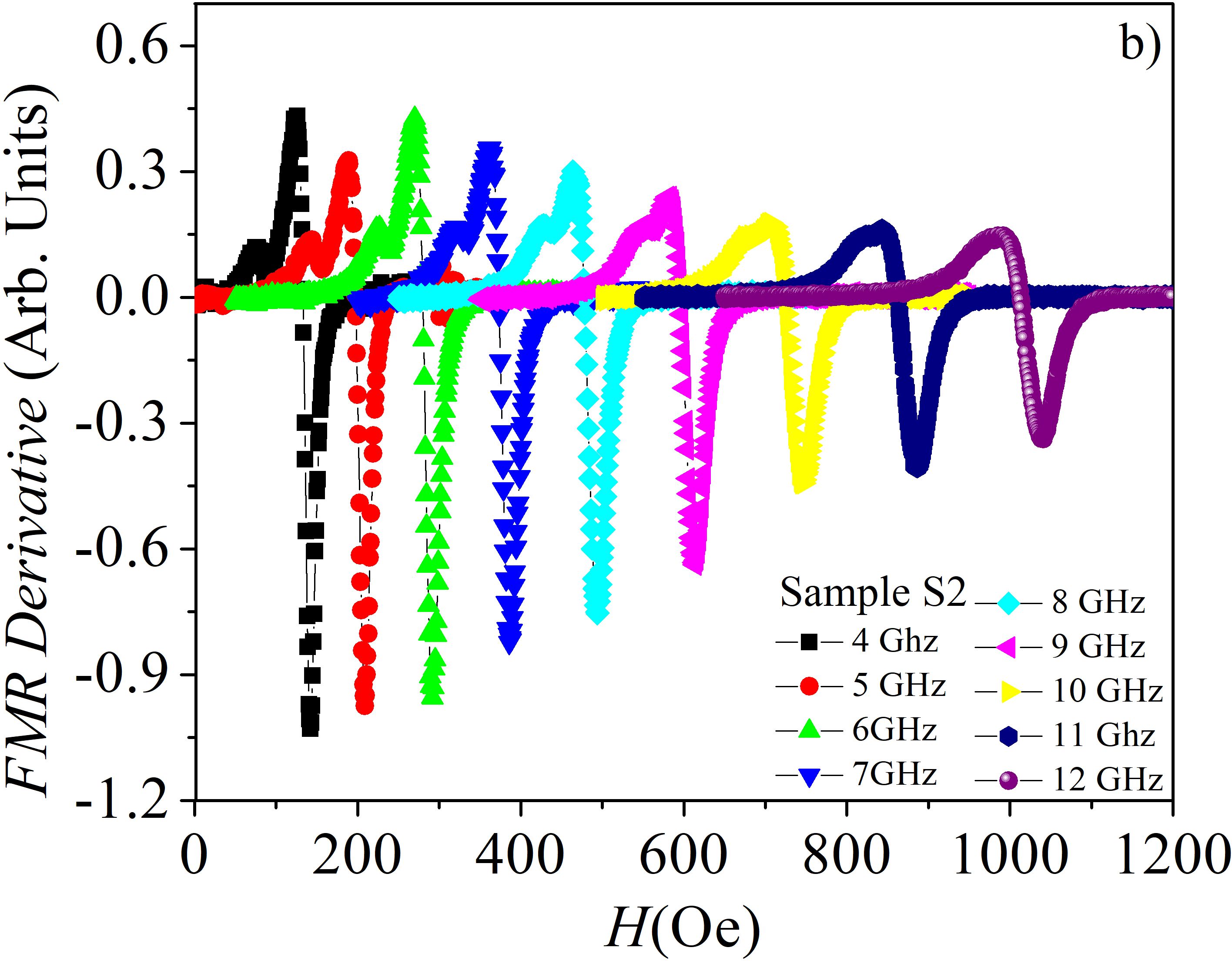}\hfill
	\includegraphics[width=.3\textwidth]{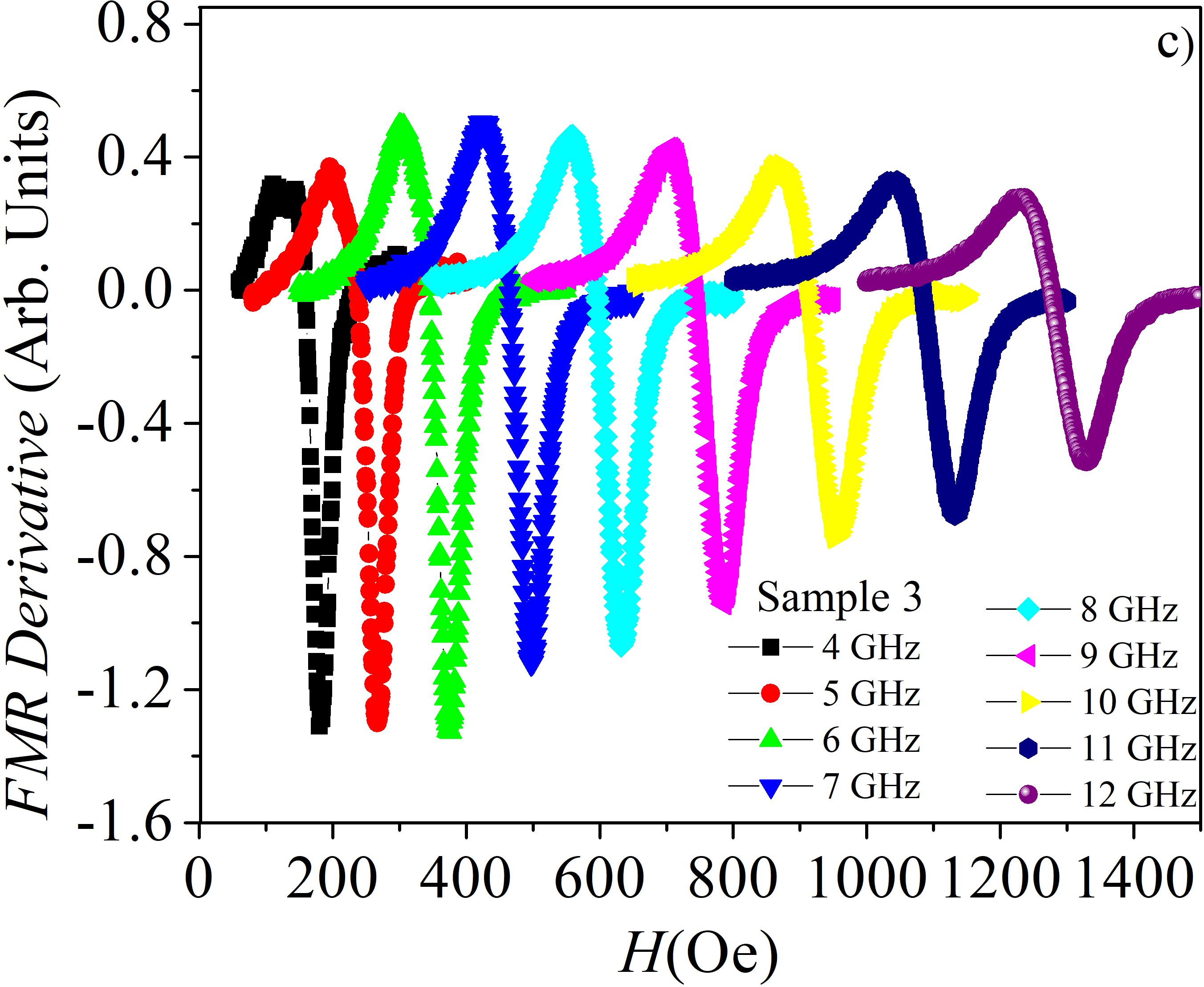}
	\caption{FMR Spectra for a) S1 b) S2 and c) S5. The r.f power used in this case is around 11 mW.}
	\label{fig:ISHE}
\end{figure*}

Figure S2 depicts the FMR spectra of the samples under consideration. Sample S1: Si/CoFeB(10), Sample S2: Si/CoFeB(10)/IrO$_2$(2), Sample S5: Si/CoFeB(10)/Ir(3)/Cr(3). The number in brackets are in nm. During experiment we have performed FMR from 4Ghz to 12 GHz frequency with intervals of 0.5 GHz. For clarity purposes, spectra at interval of 1 GHz is shown. The spectra is fitted to a sum symmetric and anti-symmetric lorentzian function.
\begin{eqnarray}
	Signal= 4 A\frac{\Delta H(H-H_{res})}{(4(H-H_{res})^2+\Delta H^2)^2} 
	-S\frac{\Delta H^2-4(H-H_{res})^2}{(4(H-H_{res})^2+\Delta H^2)^2} \\\notag +offset + slope.H 
	\label{FMR Fitting Equation one}
\end{eqnarray}
Here \textit{H} is the applied field. $\Delta H$ is the linewidth, $H_{res}$ is the resonant field. A and S denote the weight of the anti-symmetric and symmetric part of the function respectively.

\section{Angle dependent ISHE voltage measurements}

We have also performed angle dependent ISHE voltage measurement for Si/CFB(10)/IrO$_2$(3,5) (Samples S3 and S4). The plots for all the samples are provided together for comparison.

\begin{figure}[h]
	\centering
	\includegraphics[width=1\linewidth]{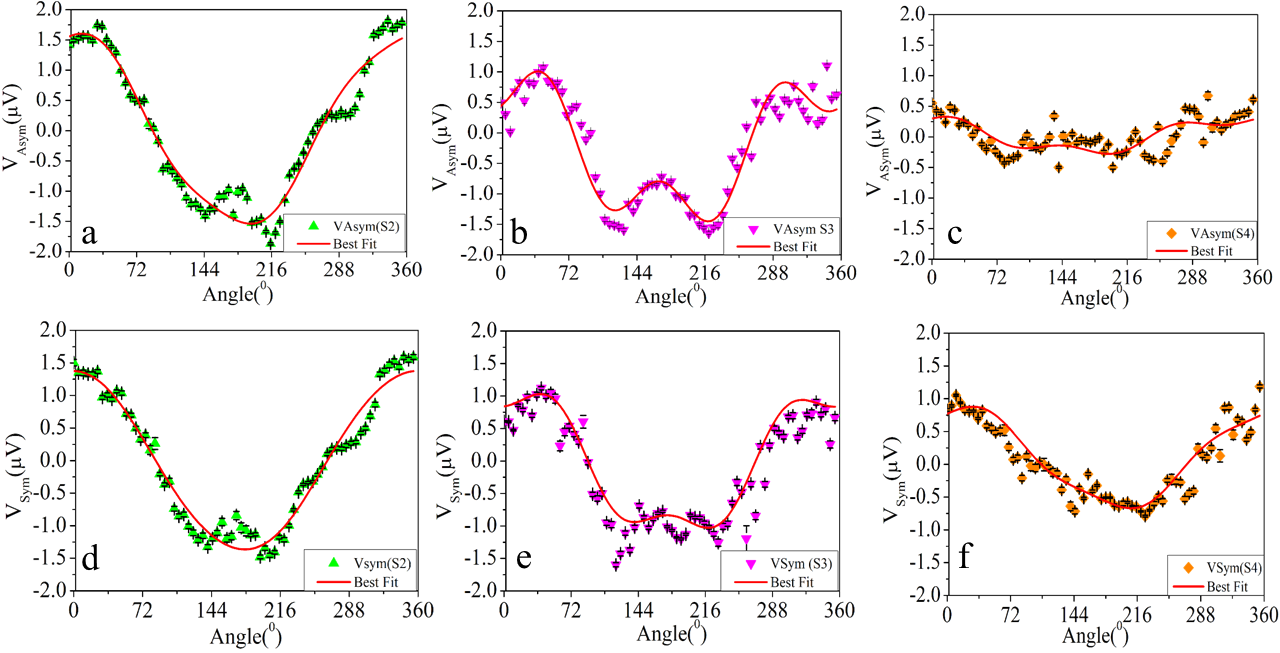}
	\caption{In-plane angular dependence of obtained voltage with respect to the external magnetic field. The plots are for the symmetric components extracted from the overall signal as shown in Figure 4 in main manuscript.  }
	\label{fig:anglesym}
\end{figure}